# Another Way to Realize Maxwell's Demon
## A Non-uniform Magnetic Field Provides a One-way Channel for Thermal Electrons

Xinyong Fu, Zitao Fu, Shanghai Jiao Tong University, September 28, 2005

Room 302, No. 7, Lane 24, South Yili Road, Shanghai 201103, P. R. China

xyfu@sjtu.edu.cn

Abstract

This is another approach to realize Maxwell's "demon" hypothesis. Two Ag-O-Cs thermal electron ejectors A and B are settled in a vacuum tube. A non-uniform magnetic field exerted on the tube provides a one-way channel for the thermal electrons. Ejector A, losing electrons, charges positively, while ejector B, getting electrons, charges negatively, resulting in an electric voltage. In flying from A to B, the speed of the electrons decreases, and part of their thermal kinetic energy converts into electric potential energy. Thus, the temperature of the whole electron tube drops down slightly, and that can be compensated by the heat attracted from the ambient air. The device can provide continuously a small but macroscopic power to an external load, violating Kelvin's statement of the second law.

Text

In a long time past, one of the authors of this paper has designed and executed successfully an experiment in which thermal electrons ejected by two ejectors are so controlled by a uniform magnetic field that they can fly only from one ejector to the other, as shown in Fig. 1, and part of the thermal kinetic energy of the transiting electrons converts into electric potential energy [1][2]. The event converts all the heat attracted by the vacuum tube from the ambient air, which is at a single and stable room temperature, into electric energy, without producing any other effect. It violates Kelvin's statement of the second law of thermodynamics, realizing the famous "demon" hypothesis put forward by James Clerk Maxwell in 1871 [3][4].

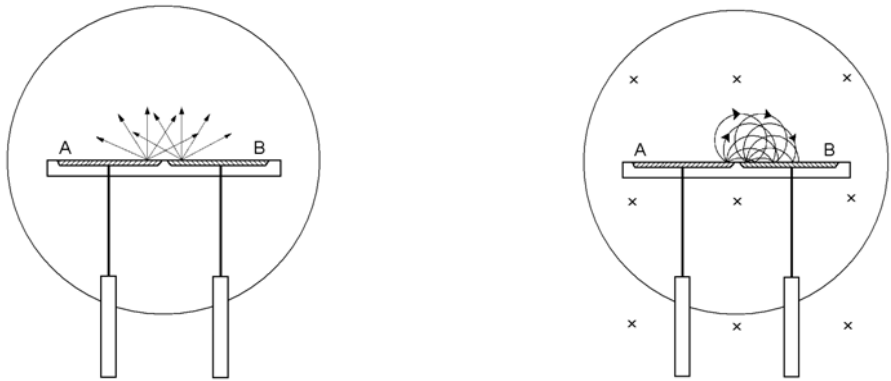

(a) The motion of thermal electrons when there is no magnetic field

(b) A uniform magnetic field controls the motion of thermal electrons

Fig. 1 Maxwell's demon (a uniform magnetic field) controls thermal electrons' motion.

Recently, the authors have found another method to realize the famous hypothesis. A non-uniform magnetic field can also provide a one-way channel for thermal electrons ejected by two ejectors, playing the role of a Maxwell's demon. The two methods are of different approaches but equally satisfactory results. The idea and mechanism of the new method is as follows.

Two identical plate-like thermal electron ejectors A and B are settled in a vacuum tube, as shown in Fig. 2. The ejectors are parallel and toward each other. It is rather good to choose Ag-O-Cs photoelectric cathodes or some analogous material to be the ejectors, as they eject thermal electrons with certain intensity even at room temperature [5]. With such ejectors, we can let the whole device work at a single room temperature to get a simple and ideal situation, in other words, to imitate an ideal model for the test. A static non-uniform magnetic field is applied to the vacuum tube, with a magnetic induction weakening gradually from left to right.

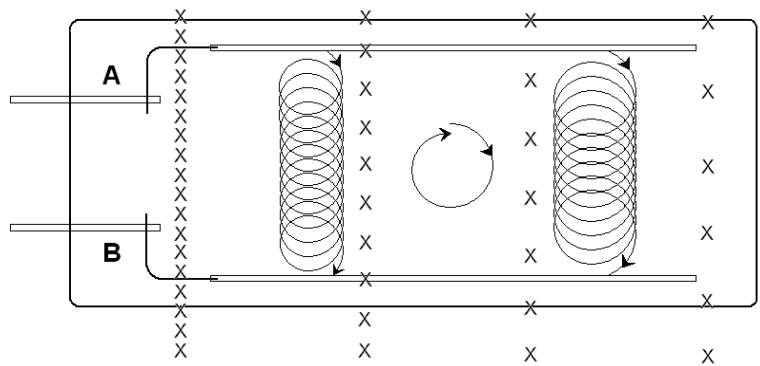

Fig. 2 The structure of the new vacuum electronic tube; the distribution of non-uniform magnetic field; and the cycling motions of thermal electrons.

First, we analyze how an electron, which moves at a certain speed, cycles in such a non-uniform magnetic field. Suppose the electron starts its motion with a rightward velocity. A trace of a whole-cycle of the electron is shown in the central part of the tube in Fig. 2: the initial velocity is rightward, and after a whole-cycle, the velocity becomes rightward again. For convenience of analysis, imagine separating the whole-cycle into two halves, the right half and the left half. For the right half, the average magnetic field is weaker, so its average radius of curvature is bigger, and hence the "diameter" of this half is longer. Contrary, for the left half, the average magnetic field is more intense, so its average radius of curvature is smaller, and hence the "diameter" of this half is shorter. Therefore, in a whole-cycle, the electron drifts downward a small displacement δ. After the first whole-cycle, the electron continues to cycle repeatedly, drifting downward cycle by cycle until it meets ejector B and gets absorbed. (The situation is rather similar to an electron's cycloid motion in a magneto-electric orthogonal field.)

Now let's look at whether an electron ejected by A or by B will return back to its original ejector (or mother ejector) directly, or will it cycle away from its original ejector toward the other ejector.

We first check up on the motions of thermal electrons ejected by A, the top ejector. Fig. 3 shows five electrons ejected by A at a point O in five different directions.

An electron ejected by A in the direction of $OA_1$ (rightward), moves firstly along a whole-cycle trace, namely, arc $OSA_1'$, as shown in Fig. 4(a). In the duration, the electron drifts down a displacement δ. So, it does not return back to A. Instead, after passing a new top point $A_1'$, it starts the second cycle, then the third cycle, and so on, and moves downward cycle by cycle until it touches the other ejector B and gets absorbed.

An electron ejected by A in the direction of $OA_2$ moves along arc $OSA_2'$, as shown in Fig. 4 (b). It returns back finally to A. Apparently, it does not go to ejector B.

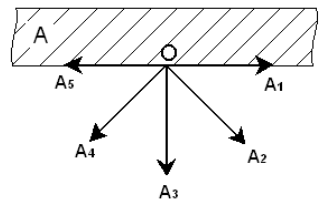

Fig. 3 Electrons ejected by A at point O in five different directions.

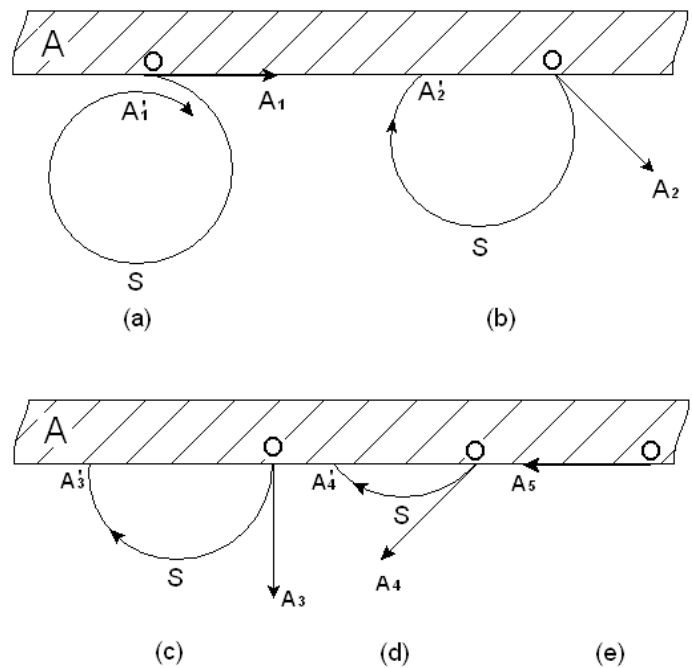

Fig. 4 The individual traces of electrons ejected by A in the five different directions.

An electron ejected by A in the direction of $OA_3$ moves along arc $OSA_3'$, as shown in Fig. 4(c). It returns back to A in about half a cycle.

An electron ejected by A in the direction of $OA_4$ moves along arc $OSA'_4$, as shown in Fig. 4(d). It returns back to A quickly.

An electron ejected by A in the direction of $OA_5$ cannot leave A, as it is ejected leftward. It just stays within the ejector, as shown in Fig. 4(e).

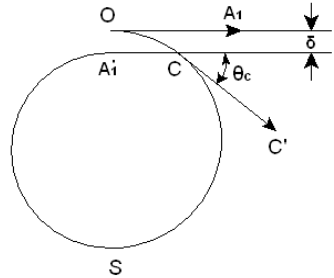

Fig. 5 How many electrons ejected by A will leave A and go to B?
All the electrons whose ejection angle is smaller than $\theta_c$.

Now, an electron ejected in the direction of $OA_1$ can leave A and cycle toward B. Similarly, a part of the electrons ejected in directions close to $OA_1$ should also be able to leave A and cycle toward B. How many are these electrons?

In order to answer the question, let's draw from point $A'_1$ in Fig. 4 (a) a straight line $A'_1C$ parallel to $OA_1$, crossing with the whole-cycle trace $OSA'_1$ at C, as shown in Fig. 5. Arrow CC' is the tangential line of the whole-cycle trace at point C, and $\theta c$ is the angle between this tangential line and the horizon. It is apparent that an electron ejected by ejector A in the direction of CC′ is critical to be able to leave A and cycle toward B, since when it arrives the new top point $A'_1$ in its first cycle, it is in the same altitude as its starting point. It can just pass the top point $A'_1$ and continues to cycle further.

Hence, an electron ejected in any direction between $OA_1$ and CC′, i.e., with an ejection angle between 0 and $\theta c$, does not return back to A. After reaching a new top point $A'_1$, it cycles continuously in the non-uniform magnetic field and drifts downward to B.

(In Fig. 2, we have shown two traces of such electrons, each start from ejector A, cycle repeatedly to cross the space between A and B, and finally get absorbed by B.)

And on the other hand, an electron ejected in any direction between CC′ and $OA_5$, i.e., with an ejection angle between $\theta c$ and 180°, such as the electrons ejected in the directions of $OA_2$, $OA_3$ and $OA_4$, which we have discussed above, returns back to ejector A along a trace less than a whole-cycle.

If we increase the gradient of the magnetic field, the distance between $OA_1$ and $A'_1C$ (i.e., δ) will increase, too. Then the critical ejection angle $\theta c$ will increase, and more electrons leave A and move toward B, and fewer electrons ejected by A return back to A directly.

(In Fig. 3 and Fig. 4, the thermal electron ejecting surface of A is flat. If there are appropriate small and parallel dents or ditches on the surface, more electrons will be able to leave ejector A when the tube works.)

Now let us check up on the motions of electrons ejected by B, the bottom ejector. Fig. 6 shows five electrons ejected by B at a point O in five different directions.

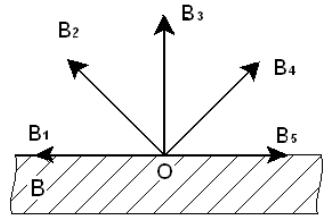

Fig. 6 Electrons ejected by B at a point O in five different directions.

An electron ejected by B in the direction of $OB_1$ (leftward) should move along a trace arc $OTB_1'$, as shown in Fig. 7 (a). It returns back to B after a travel just a little less than a whole-cycle. Therefore, it cannot pass through the space between B and A to reach the top ejector A.

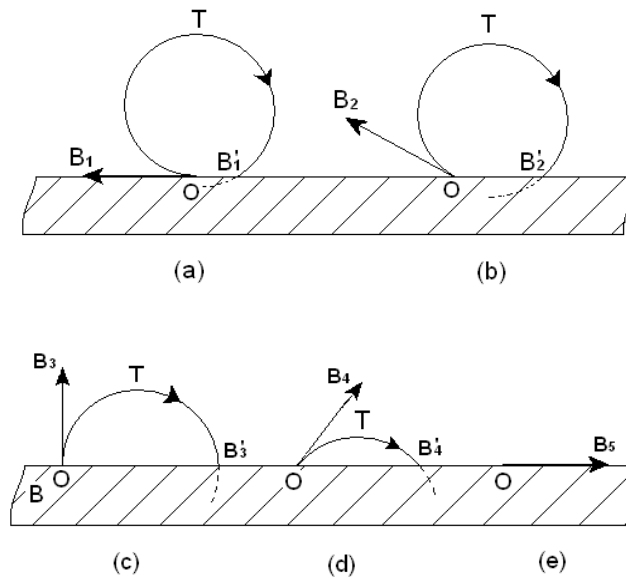

Fig. 7 The individual traces of electrons ejected by B in the five different directions. All the electrons ejected by B return back to B directly.

Electrons ejected by B in the directions of $OB_2$, $OB_3$ and $OB_4$ should move along arcs $OTB_2'$, $OTB_3'$ and $OTB_4'$, respectively, as shown in Fig. 7 (b), (c) and (d). All these electrons return back to B directly along traces less than a whole-cycle.

An electron ejected by B in the direction of $OB_5$ does not leave B, as its initial direction is rightward. It just stays within B, as shown in Fig. 7(e).

In a word, all the electrons ejected out by B would return back to B directly in less than a whole-cycle, no matter in what a direction they are ejected. None of the electrons ejected by B would cross the space between B and A to reach the top ejector A.

To sum up, part of the electrons ejected by A can cycle and move downward to B, but none of the electrons ejected by B can move upward to reach A. Thus, the non-uniform magnetic field between A and B provides a one-way channel for the thermal electrons. The authors hold that such a one-way channel is obviously another Maxwell's demon dealing with thermal electrons.

In this way, ejector A charges positively as it loses electrons, while ejector B charges negatively as it gets electrons. Such a charge distribution means an electric voltage between A and B; the electric potential energy of an electron at B is now higher than the one at A. The vacuum tube now becomes a power source. Connecting an appropriate load to the lead-out wires of A and B, the vacuum tube can provide a small but macroscopic power to the load continuously.

Where does the electric energy of the power source come from?

It is easy to answer the question. As an electron cycles and drifts downward from A to B, its electric potential energy increases cycle by cycle, and hence its thermal kinetic energy (and its speed) must decrease cycle by cycle. How much electric potential energy it gains equals exactly how much thermal kinetic energy it loses. (In Fig. 2, in the two continuous cycling curves, the average radius for every whole-cycle should decrease cycle by cycle as the electrons drift down. Nevertheless, for simplicity in drawing, we didn't show such details in the figure.) Of course, during the transition from A to B, the "temperature" of the transiting "electron gas" drops down, and this drop down then results in a slight drop down of the temperature of the whole vacuum tube (the heat capacity of the vacuum tube is much greater than the transiting "electron gas".) The drop down of the temperature of the vacuum tube can be compensated by the heat attracted by the vacuum tube from the ambient air in the laboratory. We may suppose that the air in the laboratory is at a stable room temperature.

In such a process, heat is attracted from a single reservoir (the ambient air) and converted totally into work without producing any other effect, in contradiction to Kelvin's statement of the second law. It shows again how ingenious and elegant Maxwell's 1871 hypothesis was!

## Discussion

Some people may ask, electrons "vaporize" at A and then "condense" at B, resulting in A's cooling down and B's warming up. Will this be an obstruction to the experiment?

This is not an obstruction to the experiment.

The first reason is, the difference in temperature produced in this way is generally very slight, which would not affect the above-mentioned process.

Second, a more important reason is, from the viewpoint of thermodynamics, such a conversion from a uniform temperature to a temperature difference means a decrease in entropy. So, it means that things have become better, or we have got some advantage. If we think it is trivial to use such a tiny benefit or advantage, we may choose to abandon it. As is well known, it is very easy to abandon such a benefit or advantage: take a thermal relaxation, for example, a spontaneous process of heat radiation and conduction, the difference in temperature will disappear by itself in a short period.

Nevertheless, we may choose not to abandon the benefit or advantage as well, and in so doing, we might meet some interesting phenomena.

First, let us short-circuit the external load (that means the current becomes greater.) Now, the electrons, after cycling and moving from A to B within the tube, go back from B to A immediately through the external short-circuit. Thus, relying on its own thermal motion, and with the help given gratis by the non-uniform magnetic field, the electrons flow ceaselessly in one direction in the closed circuit. Note that it is a macroscopic direct current, and it can maintain by itself forever, just like the thermal random motion of the molecules in a gas at an equilibrium state can maintain forever by itself!

Furthermore, as mentioned above, the electrons cycling in the circuit take away "evaporation heat" from ejector A, and give "condensation heat" to ejector B. The temperature of A drops down as it gives heat off, while the temperature of B rises up as it takes heat in. We will thus, without expenditure of work, raise the temperature of B and lower that of A.

A gratis conversion from a uniform temperature to a difference in temperature was exactly the bold and excellent hypothesis (as well as a beautiful dream) of James Clerk Maxwell, which he described in detail in his textbook *Theory of Heat* published in 1871.

A few further word. The work function of Ag-O-Cs cathode today is about 0.7eV. Is it possible for mankind to get in future a cathode material with a work function of 0.3eV? Or even a cathode material with a work function of 0.1eV? If that is possible, then, with such new cathode material, and together with some other technologies (for example, large-scale integrating), it may be hopeful to put the two Maxwell's demons (uniform magnetic field and non-uniform magnetic field) into practical uses, both for energy conversion and for producing difference in temperature.